\documentclass{nature}

\usepackage{graphicx}
\usepackage{amsmath}
\usepackage{url}
\usepackage{comment}
\usepackage{chngcntr}
\usepackage{multibib}
	
\def\apj{\it Astrophys. J.}                 
\def\apjl{Astrophys. J.}

\def\aap{\it Astron. Astrophys.}

\def\mnras{\it Mon. Not. R. Astron. Soc.}

\def\nat{Nature}

\title{Aligned Grains and Scattered Light Found in Gaps of Planet-Forming Disk}

\author{Ian W. Stephens$^{1}$, Zhe-Yu Daniel Lin$^{2}$, Manuel Fern\'{a}ndez-L\'{o}pez$^{3}$, Zhi-Yun Li$^2$, Leslie W. Looney$^4$, Haifeng Yang$^5$, Rachel Harrison$^{4,6}$, Akimasa Kataoka$^7$, Carlos Carrasco-Gonzalez$^8$, Satoshi Okuzumi$^9$, Ryo Tazaki$^{10,11}$}

\begin{document}

\newcommand\arcsec{\mbox{$^{\prime\prime}$}}%

\maketitle

\begin{affiliations}
 \item Department of Earth, Environment, and Physics, Worcester State University, Worcester, MA 01602, USA
\item Astronomy Department, University of Virginia, Charlottesville, VA 22904, USA
 \item Instituto Argentino de Radioastronom{\'i}a, CCT-La Plata (CONICET), C.C.5, 1894, Villa Elisa, Argentina
\item Department of Astronomy, University of Illinois, Urbana, Illinois 61801, USA
 \item Kavli Institute for Astronomy and Astrophysics, Peking University, Yi He Yuan Lu 5, Haidian Qu, Beijing 100871, People’s Republic of China
 \item School of Physics and Astronomy, Monash University, Clayton VIC 3800, Australia
\item National Astronomical Observatory of Japan, Osawa 2-21-1, Mitaka, Tokyo 181-8588, Japan 
 \item Instituto de Radioastronomía y Astrof{\'i}sica (IRyA), Universidad Nacional Autónoma de México (UNAM), Mexico 
 \item Department of Earth and Planetary Sciences, Tokyo Institute of Technology, Meguro, Tokyo 152-8551, Japan 
 \item Universit\'{e} Grenoble Alpes, CNRS, Institut de Plan\'{e}tologie et d'Astrophysique (IPAG), F-38000 Grenoble, France 
 \item Astronomical Institute, Tohoku University, Sendai 980-8578, Japan 
\end{affiliations}
\begin{abstract}

Polarized (sub)millimeter emission from dust grains in circumstellar disks was initially thought to be due to grains aligned with the magnetic field\cite{Rao2014,Stephens2014b}. However, higher resolution multi-wavelength observations\cite{Kataoka2017,Stephens2017c,Harrison2019} along with improved models\cite{Kataoka2015,Yang2016a,Yang2021,Lin2022MNRAS.512.3922L,Lin2023MNRAS.520.1210L} found that this polarization is dominated by self-scattering at shorter wavelengths (e.g., 870\,$\mu$m) and by grains aligned with something other than magnetic fields at longer wavelengths (e.g., 3\,mm). Nevertheless, the polarization signal is expected to depend on the underlying substructure\cite{Pohl2016,Ohashi2019ApJ...886..103O,Lin2020b}, and observations hitherto have been unable to resolve polarization in multiple rings and gaps. HL Tau, a protoplanetary disk located 147.3\,$\pm$\,0.5\,pc away\cite{Galli2018}, is the brightest Class I or Class II disk at millimeter/submillimeter wavelengths. Here we show deep, high-resolution 870\,$\mu$m polarization observations of HL Tau, resolving polarization in both the rings and gaps. We find that the gaps have polarization angles with a significant azimuthal component and a higher polarization fraction than the rings. Our models show that the disk polarization is due to both scattering and emission from aligned effectively prolate grains. The intrinsic polarization of aligned dust grains is likely over 10\%, which is much higher than what was expected in low resolution observations ($\sim$1\%). Asymmetries and dust features are revealed in the polarization observations that are not seen in non-polarimetric observations.
\end{abstract}

\begin{figure} 
\begin{center}
\hbox{\hspace{-5ex}
\includegraphics[width=1\textwidth]{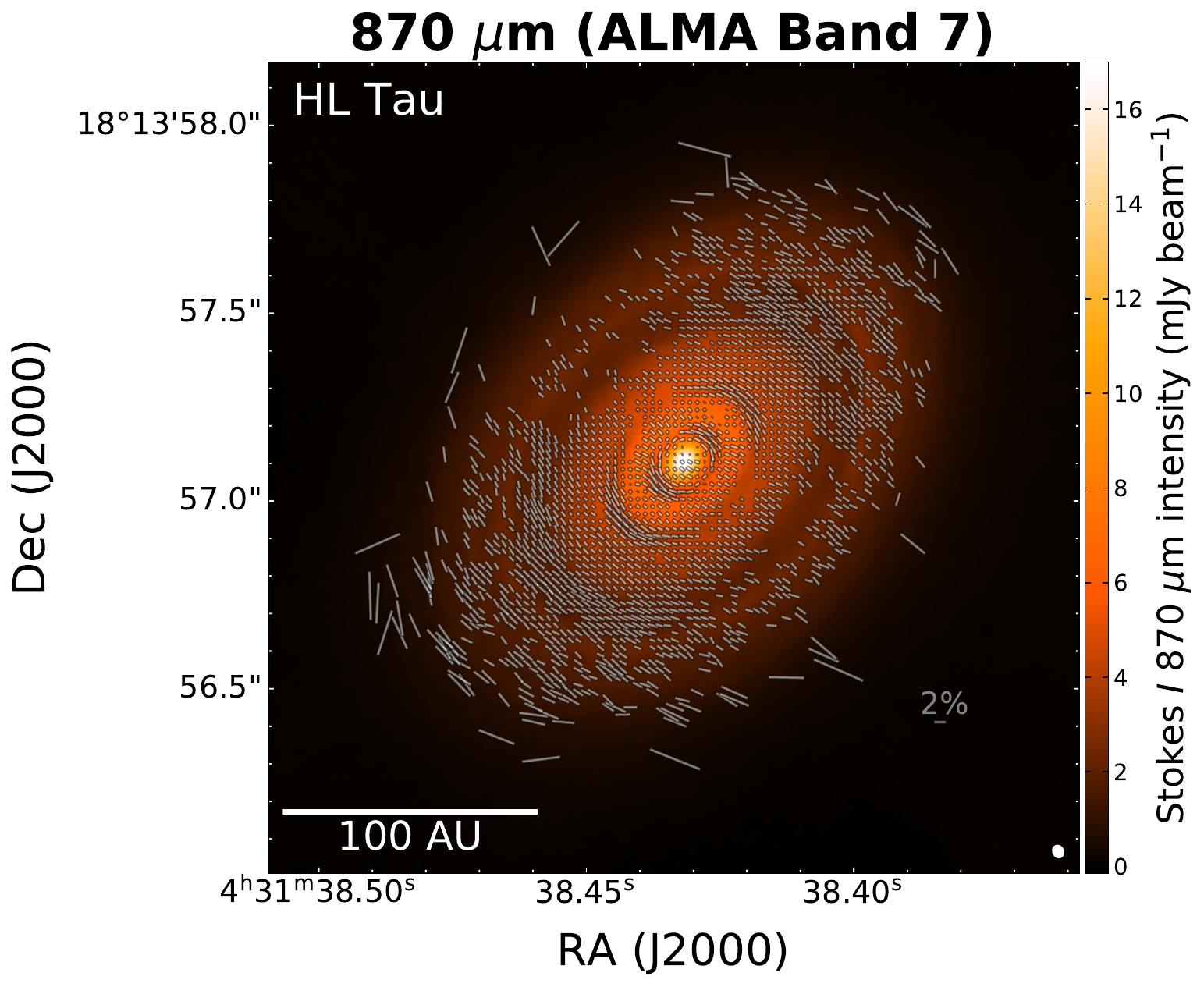}
}
\caption{ {\bf 870 $\mu$m polarization morphology overlaid on the total intensity images of HL Tau.} The length of the vectors is proportional to the percent of light that is polarized ($P$), with a 2\% scale bar shown toward the bottom right. Vectors are plotted when both their total intensity (Stokes~$I$) and the polarized intensity are detected at a $3\sigma$ level, and for $P<10\%$. Vectors with $P < 0.5\%$ are drawn as thinner lines. The resolution (beam) is shown as a small blue ellipse in the bottom right. 
}\label{fig:polmorph}
\end{center}
\end{figure}

We used the Atacama Large Millimeter/submillimeter Array (ALMA) in Full Stokes mode at 0.033$\arcsec$ (4.9\,au) resolution to observe both the total emission and polarized emission from HL Tau (see details in the Methods section). The linear polarization data (Stokes~$Q$ and $U$) were combined with data from previous observations\cite{Stephens2017c} for increased sensitivity, which resulted in a slightly coarser resolution of $\sim$0.0345$\arcsec$ (5.1\,au). In Figure~\ref{fig:polmorph}, we show the polarization morphology of HL Tau overlaid on the Stokes~$I$ image. Vectors are plotted in a grid with a separation of 0.0198$\arcsec$ (i.e., sampled slightly less than Nyquist sampling), for a total of 2067 displayed vectors. The amount of independent vectors detected toward HL Tau is well over an order magnitude more than detected previously\cite{Stephens2017c}, and about an order of magnitude more vectors than the previously best resolved disk in polarization\cite{Ohashi2018}.

\begin{figure} 
\begin{center}
\includegraphics[width=1\textwidth]{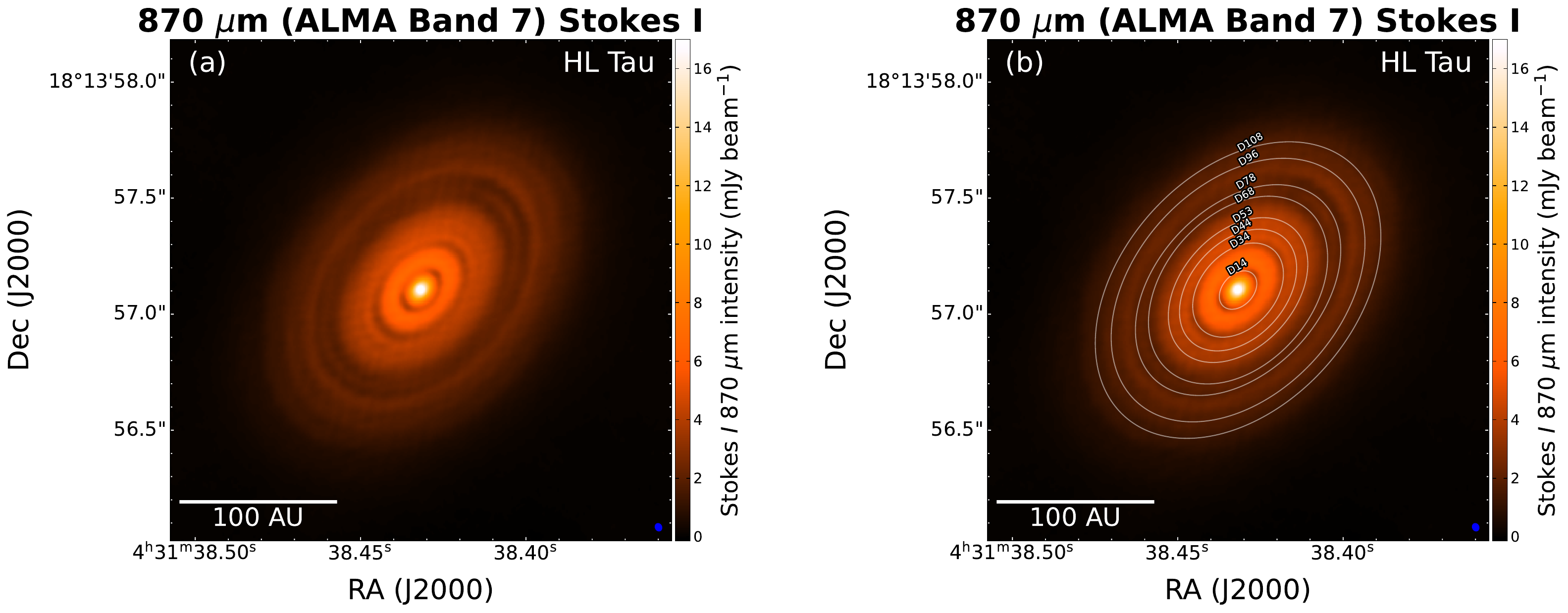}
\caption{ {\bf 870 $\mu$m total intensity (Stokes~$I$) image of HL Tau.} The resolution (beam) is shown as a small blue ellipse in the bottom right. Panel a and b are identical, except panel b has ellipses at the locations of the gaps\cite{ALMA2015}, which includes the previously unidentified eighth gap. The number in each gap's label indicates the semimajor axis length in au.
}\label{fig:StokesI}
\end{center}
\end{figure}


Given the long integration times, HL Tau's Stokes~$I$ image is at much higher fidelity than what has been done before\cite{ALMA2015}, and it is presented in Figure~\ref{fig:StokesI}. Previously HL Tau had 7 identified dark bands\cite{ALMA2015} or gaps, but our new observations clearly reveal an eighth gap toward the edge of the disk, which we measure to be at a radius of $\sim$0.73$\arcsec$ or 107.5\,au. We rename the gaps from their old nomenclature\cite{ALMA2015} (D1-D7) to a nomenclature that shows their distance in au, as the 147.3\,pc distance has now been accurately determined by Gaia\cite{Galli2018}. 

\begin{figure} 
\begin{center}
\includegraphics[width=1\textwidth]{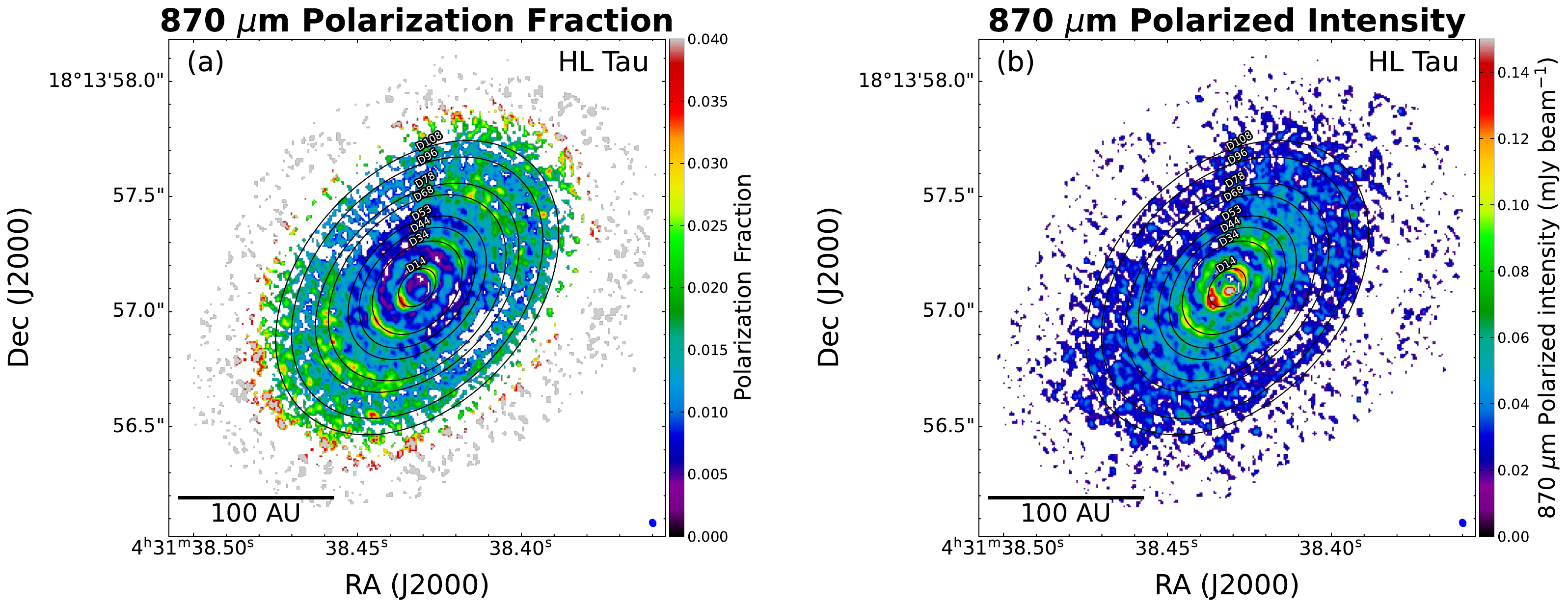}
\caption{ {\bf Polarization fraction and intensity maps of HL Tau.}  The gaps in the disk from Figure~\ref{fig:StokesI} are shown. The resolution (beam) is shown as a small blue ellipse in the bottom right. 
}\label{fig:polimages}
\end{center}
\end{figure}

Previous observations of HL Tau\cite{Stephens2017c} showed polarization vectors close to uniform and parallel to the minor axis of the disk with polarization fractions of $\sim$1\%. Figure~\ref{fig:polmorph} reveals a non-uniform polarization morphology and higher polarization fractions. The orientations of the polarization vectors in the rings are mostly uniform, while the vectors in the gaps have a strong azimuthal component, showing two distinct brackets around the disk center. Figure~\ref{fig:polimages} shows the polarization fraction and polarized intensities along with the eight gaps overlaid on HL Tau. Along with  what is seen in Figure~\ref{fig:polmorph}, there are three key features of the polarization:
\begin{enumerate}
    \item Polarization fractions are typically much higher in the gaps (reaching up to 3.7\%) than in the rings. Even the polarized intensity is frequently higher in the gaps.
    \item Polarization vectors are usually uniformly aligned along the minor axis, but there is a significant azimuthal component in the gaps. Notably, the first gap (D14) has a ring of polarized intensity with azimuthally oriented polarization directions that completely surrounds the inner disk. 
    \item The polarized fractions and intensities are larger along the major axis than the minor axis.
\end{enumerate}

\begin{figure} 
\begin{center}
\hbox{\hspace{-5ex}
\includegraphics[width=1\textwidth]{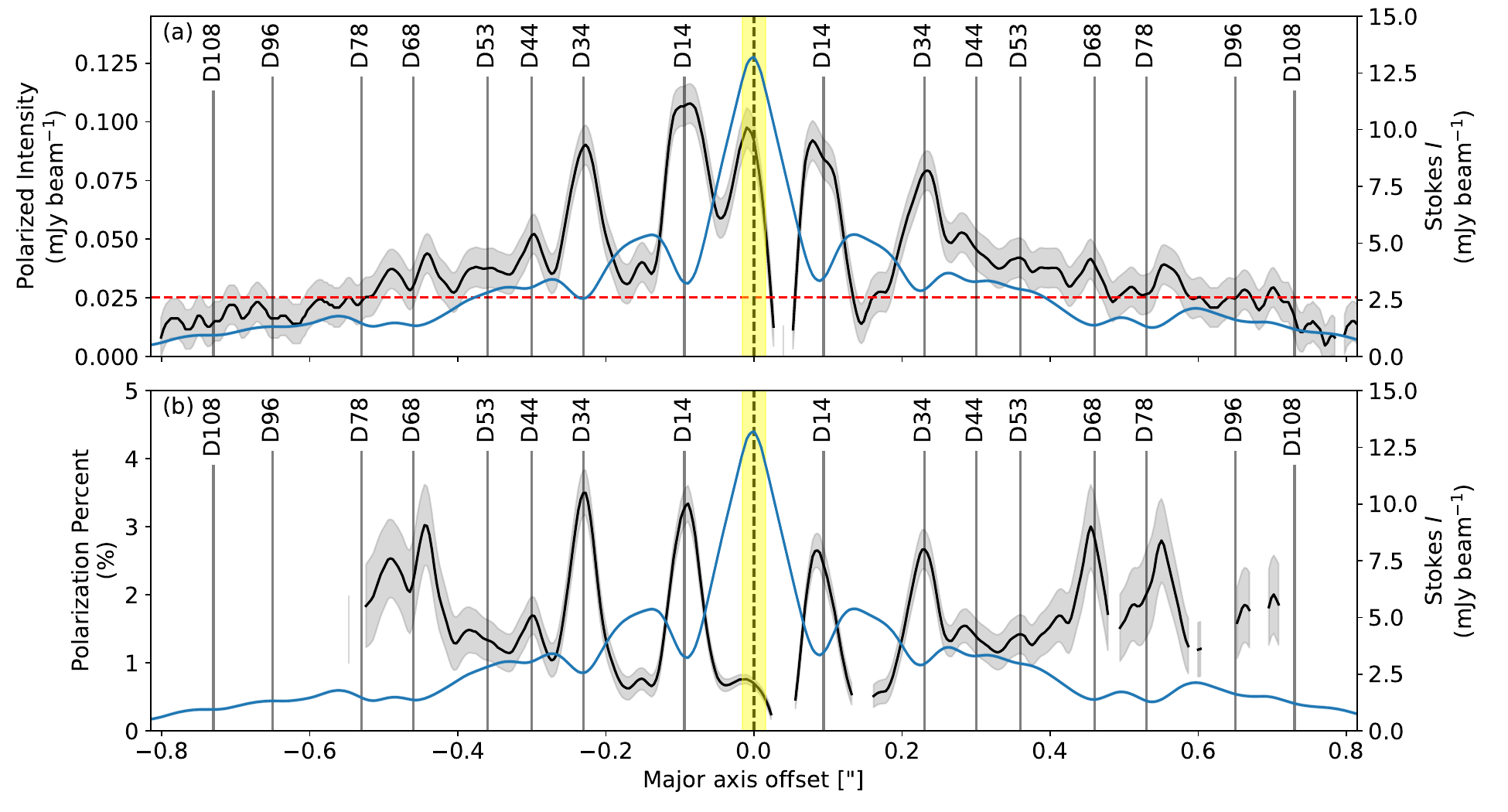}
}
\caption{ {\bf Stokes~$I$, polarized intensity , and polarization fraction profiles along HL Tau's major axis (southeast to northwest).} The Stokes~$I$ cut is in blue for both panels, while the polarized intensity and fraction is in black for panels a and b, respectively. The shaded gray shows the standard deviation (1$\sigma$). The shaded yellow shows the resolution of the observations along the major axis (0.032$\arcsec$). The red-line in the panel a is the 3$\sigma$ cut-off for polarized intensity. Panel b only shows polarization fractions measured above the 3$\sigma$ level, which is the reason why the black curve does not extend to the outer edge of the disk. Gaps are labeled with vertical lines. 
}\label{fig:cuts}
\end{center}
\end{figure}

The first feature is further quantified in Figure~\ref{fig:cuts}, which shows cuts in Stokes~$I$, polarization fraction, and polarized intensity along the major axis. For the inner 3 gaps, there are obvious local peaks in both polarized intensity and fraction. On the other hand, for Stokes~$I$ rings, there are local troughs, with the exception of the very central Stokes~$I$ peak.

The fact that the polarized intensity is stronger in the gaps was unexpected, as there is less dust. Furthermore, the Stokes~$I$ is typically more symmetric along the major axis than the polarized intensity. For example, the Stokes~$I$ fluxes along the cut for the first gap (D14) differ by 6\%, while the polarized intensity and fraction differ by $\sim$30\%. An increased asymmetry of these quantities is not expected along the major axis of an axisymmetric disk for polarization due to scattering or grain alignment, suggesting there are features in the disk not seen in Stokes~$I$. For example, there could be a slight warp in the disk, a difference along the major axis in the dust scale height, or difference in dust grain properties (e.g., composition, albedo, or sizes)\cite{Kataoka2015}.

On the other hand, for an inclined dust disk that is not geometrically thin, asymmetries are fully expected along the minor axis, with the near side of the disk (i.e., the part of the disk closer to the observer) expected to be more polarized than the far side of the disk\cite{Yang2017}. The near side of HL Tau is toward the bottom right\cite{ALMA2015}, yet the ring between the first two gaps clearly has more polarized intensity toward the far-side of the disk. This again suggests that there are some asymmetries along the minor axis. However, interior to the first gap (i.e., the central bright emission), the polarized intensity is higher on the near side compared to that on the far side, as expected for a thicker dust disk. 

Building upon previous work\cite{Lin2022MNRAS.512.3922L}, we create a model that can broadly explain the three key features mentioned earlier, with the details of the model in the Methods section. Our first attempts of a model with polarization solely due to scattering of spherical grains, which is also applicable to randomly aligned grains, could mostly reproduce the three key features. However, it cannot recreate the large polarized intensity with an azimuthally oriented polarization morphology\cite{Lin2020b} in the first gap (D14). By including both polarized thermal emission and scattering of aligned effectively prolate grains spinning about its long axis, the model not only explains all three features, but fits them more closely than the scattering-only model. In our model, we find that the long axis of a grain is aligned azimuthally around the disk which is also consistent with the polarization morphology at 3.1\,mm (Ref. \citenum{Yang2019}). The effective sizes of the prolate grains follow an $a^{-3.5}$ power-law distribution\cite{Mathis1977} with a maximum grain size of 100\,$\mu$m and creates the general scattering polarization morphology. Prolate-like grains are favored over oblate-like grains because oblate grains have difficulty producing the observed polarized morphology and intensity along the minor axis observed in D14. Nevertheless, there is likely a mix of grain shapes, and our model suggests that the observations are most consistent with effective shapes that are predominantly prolate.

This model naturally explains why the polarization is higher in the gaps than in the rings (key feature \#1) because the rings have higher optical depth which minimizes polarization due to grain alignment because of a higher dichroic extinction. In turn, this also allows for an azimuthal component only in the gaps (key feature \#2) because the optical depth in the rings is too high to produce significant polarized emission from grain alignment, but still sufficiently low enough in the gaps that the azimuthal pattern from aligned grains dominates the uniform pattern from scattering. The model also produces larger polarization fractions and intensities along the major axis than the minor axis (key feature \#3) because the polarization due to grain alignment and inclination-induced scattering\cite{Yang2016a} are constructive along the major axis while destructive along the minor axis. 

To match the model's polarization fraction to that seen in the observations, we can adjust the intrinsic polarization of dust grain emission by changing the elongation of the grain. Assuming the grains are perfectly aligned, we find that an intrinsic polarization is between 10\% and 15\% (prolate grain short to long axis aspect ratio of $\sim$0.85) is needed to match observations. Even though the observations show typical polarization levels of 2\% to 3\% in the gaps, the model suggests they are being smoothed (beamed-average) by the low polarization levels of the rings. As such, higher resolution observations are likely to reveal much higher polarization fractions. The high intrinsic polarization is similar to what has been found in the interstellar medium\cite{Planck19,Planck20}, even though grains in the disk have grown to a much larger size. The intrinsic aspect ratio is likely even smaller than 0.85 since that value assumed 100\% alignment efficiency.

Furthermore, the model shows that rings are optically thick and must be a factor of $\sim$10 stronger in the gaps, which has been suggested in multi-wavelength studies of HL Tau\cite{Pinte2016}. We are able to come to this conclusion just with polarization observations at a single wavelength. Since the rings are much more optically thick, most of the mass must be in these optically thick rings. These rings are thus the likely reservoirs for the future formation of planetesimals.

Grains of $\sim$100\,$\mu$m size are likely not a unique solution and would underestimate the intensity of the image at longer wavelengths (e.g., 3\,mm). Models could be further fine-tuned, but they are degenerate due to several different parameters, including grain size, composition, albedo, porosity, alignment efficiency, and shape\cite{Tazaki2019ApJ...885...52T,YangLi2020,Lin2023MNRAS.520.1210L}. Multi-wavelength high resolution observations are necessary to overcome these degenerencies and accurately constrain grain sizes. Nevertheless, these models show that both grain scattering and alignment are at play, with scattering dominating the polarization in the rings and aligned grains in the gaps. Finally, it is worth mentioning that even these models cannot explain polarization asymmetries across the major and minor axis, indicating that polarization may reveal dust properties across a planet-forming disk that cannot be seen by non-polarimetric continuum observations.





\begin{addendum}
 \item This research made use of APLpy, an open-source plotting package for Python\cite{Robitaille2012}. This paper makes use of the following ALMA data: ADS/JAO.ALMA\#2016.1.00115.S and ADS/JAO.ALMA\#2019.1.01051.S. ALMA is a partnership of ESO (representing its member states), NSF (USA) and NINS (Japan), together with NRC (Canada), MOST and ASIAA (Taiwan), and KASI (Republic of Korea), in cooperation with the Republic of Chile. The Joint ALMA Observatory is operated by ESO, AUI/NRAO and NAOJ. The National Radio Astronomy Observatory is a facility of the National Science Foundation operated under cooperative agreement by Associated Universities, Inc. Z.-Y.D.L. acknowledges support from NASA 80NSSC18K1095, the Jefferson Scholars Foundation, the NRAO ALMA Student Observing Support (SOS) SOSPA8-003, the Achievements Rewards for College Scientists (ARCS) Foundation Washington Chapter, the Virginia Space Grant Consortium (VSGC), and UVA research computing (RIVANNA). Z.-Y.L. is supported in part by NASA 80NSSC20K0533 and NSF AST-2307199. L.W.L and R.H. acknowledges support from NSF AST-1910364 and NSF AST-2307844. C.C.-G. acknowledges support from UNAM DGAPA-PAPIIT grant IG101321 and from CONACyT Ciencia de Frontera project ID 86372. 
\item[Author Contributions] Project was led by I. Stephens. Polarization modeling was performed by Z.-Y. Lin. Data reduction was performed by M. Fern\'{a}ndez-L\'{o}pez.  All authors analyzed and discussed the observations and manuscript.
\item[Author Information] Reprints and permissions information is available at www.nature.com/reprints. The authors declare no competing financial interests. Readers are welcome to comment on the online version of the paper. Correspondence and requests for materials should be addressed to I. Stephens (istephens@worcester.edu).
\item[Data Availability] The observational data products generated and analyzed during the current study are available in the dataverse repository \url{https://dataverse.harvard.edu/dataverse/HLTau_Band7_Pol} under DOIs \url{https://doi.org/10.7910/DVN/MM4V5M} and \url{https://doi.org/10.7910/DVN/7CQRBC}. Raw ALMA data are available at \url{https://almascience.nrao.edu/aq/}. Additional datasets (e.g., modeling) are available from the corresponding author on request.
\end{addendum}

\newpage

\newpage
\begin{methods}
\subsection{Observations and Data Reduction}
Observations presented in this paper are from ALMA project 2019.1.01051.S, which was combined in part with ALMA project 2016.1.00115.S, as discussed below (PIs: I. Stephens). Project 2019.1.01051.S observed HL Tau in Full Stokes polarization at 870\,$\mu$m using two different array configurations, C-5 and C-8, which together provide baselines from 54\,k$\lambda$ to 13300\,k$\lambda$. The observations were taken from June to October 2021, with a total time on source of 18.7 hours. Each of the tracks observed used between 39 and 43 antennas. We tuned the correlator for four 1.75\,GHz spectral windows centered at sky frequencies of 336.5, 338.5, 348.5, and 350.5\,GHz. We downloaded the raw measurement sets of the observations and applied the initial calibration delivered in a pipeline by the ALMA staff. 

The Stokes~$I$ continuum emission was created using only the 2019.1.01051.S data.  They were self-calibrated using three iterative phase-only calibration and cleaning stages, with progressively lower time intervals down to 10.4 seconds. We then removed 9 channels contaminated by line emission from C$^{17}$O, SO$_2$, and CH$_3$CO in different spectral windows, which account for $\sim$3\% of the total bandwidth. Following the analysis by the Disk Substructure at High Angular Resolution Project (DSHARP) team\cite{A18}, before combining the datasets of the two different configurations, we aligned the disk peak in both datasets using CASA programs \texttt{fixplanets} and \texttt{phaseshift} and re-scaled their fluxes using \texttt{gencal}. The final Stokes~I image has an rms noise level of 15.9$\mu$Jy\,beam$^{-1}$, which gives a signal-to-noise ratio of 1073 at the peak. It was constructed using a robust weighting of 0.5, and the synthesized beam was 35.5 mas $\times$ 30.5 mas at a position angle of 22$^\circ$ (measured counterclockwise from North). We also attempted to combine these Stokes~$I$ data with HL Tau Science Verification data\cite{ALMA2015}, but we found that these data reduced the signal to noise.

For producing the polarization images, we combined data from ALMA project 2016.1.00115.S\cite{Stephens2017c}, but self-calibrating in the same manner as above, and aligning the peak position, and re-scaling the flux of the visibilities to match that of the 2019.1.01051.S combined data. This project comprised shorter baselines from the ALMA C-4 configuration, which gave baselines as short as 15.5\,k$\lambda$. The frequency setup was the same as the 2019.1.01051.S data, and we also removed the same 9 channels with line contamination. The Stokes\,$I$ image of this track had significant side-lobes, and only added additional noise to the 2019.1.01051.S Stokes $I$ data (rms noise level of 15.9$\mu$Jy\,beam$^{-1}$ without C-4 versus 39.7$\mu$Jy\,beam$^{-1}$ with C-4).  As such, for Figures~\ref{fig:polmorph} and~\ref{fig:StokesI}, we did not include the C-4 data for the Stokes~$I$, but we did for the polarization vectors. However, for the rest of the figures in the paper and extended data, we include the C-4 in Stokes~$I$, as a consistent beam and spatial sampling is needed for proper calculations (i.e., polarization fraction) and comparisons of radial cuts. After a H\"ogbom CLEAN deconvolution interactive process in CASA using a robust parameter of 0.5, the final beam size for the polarization data was 37.4 mas $\times$ 31.8 mas with a position angle of 26$^\circ$, reaching a rms noise level of 8.4$\mu$Jy\,beam$^{-1}$. We found that combining these data (2016.1.00115.S and 2019.1.01051.S) added just over 500 vectors to Figure~\ref{fig:polmorph}, with these vectors primarily corresponding to locations with low-level Stokes~$I$ emission. Since polarization fraction and intensity can only be positive, there exists a bias toward positive values. As such we de-bias all polarization data following previous works\cite{H14,H15}. As mentioned in the plots, the vectors are all detected at a 3$\sigma$ level and have $P < 10\%$. The vectors that have $P > 10\%$ are few and toward the outer disk, and thus are potentially interferometric artifacts or random noise above the 3$\sigma$ level. Including these in Figure~\ref{fig:polmorph} makes the other vectors less clear, as these vectors are long given the 2\% scale bar.

ALMA also measures Stokes~$V$, which we also imaged. Without considering instrumental effects, the Stokes~$V$ from the 2021 observational data is positive and appears to be detected significantly in some areas of the disk with a circular polarization fraction of a few tenths of a percent. However, the 2017 Band~7 polarization data\cite{Stephens2017c} are inconsistent with these observations since they suggest that the Stokes~$V$ is negative and has similarly low circular polarization fractions. Since ALMA is not currently commissioned to measure significant (3$\sigma$) polarization at levels less than 1.8\% due to instrumental effects (primarily beam squint), we consider any observational indication of circular polarization as likely from instrumental artifacts.



\subsection{Linear Polarization Modeling} \label{sec:linear_polarization_modeling}
The goal of this model is to demonstrate the main features of the observed Stokes~$I$, linear polarized intensity, and linear polarization fraction through radiation transfer modeling of scattering of aligned prolate grains using the newly developed approach of Lin et al. (2022) (Ref. \citenum{Lin2022MNRAS.512.3922L}). The model presented here is essentially identical to that presented in Lin et al. (2022) , except we consider rings and gaps within the disk rather than just a smooth density profile. Incorporating this substructure when matching the model to observations allows for new constraints on the dust parameters. In the following, we briefly describe the model and its setup, but the reader should refer to Lin et al. (2022) for more details.

Scattering-induced dust polarization depends on grain properties such as size, shape, and composition\cite{Kataoka2015, Tazaki2019ApJ...885...52T, YangLi2020, Lin2023MNRAS.520.1210L}, all of which are uncertain. Polarization data for a single wavelength do not allow a precise determination of these properties in general; our case is no exception. 
We adopt the oft-used dust properties from DSHARP\cite{B18}. Specifically, we adopt a complex refractive index of $m \approx 2.30 + 0.0228\ i$ at $\lambda=870$~$\mu$m and a power-law grain size distribution\cite{Mathis1977} of $a^{-3.5}$. From here, we fix the minimum grain size $a_{\text{min}}$ to 0.1~$\mu$m and the maximum grain size $a_{\text{max}}$ to 100~$\mu$m, and we assume the grains have no porosity. Since the scattering in the Rayleigh regime under consideration is dominated by grains containing most of the volume (and thus the mass), the value of $a_{\text{min}}$ has little effect on the model for the chosen size distribution. The value of $a_{\text{max}}$ is set by the requirement that (compact spherical) grains of order 100~$\mu$m in size are needed to dominate the scattering of 870~$\mu$m (Band 7) photons efficiently\cite{Kataoka2015}. It is well-known that this size is in tension with the larger (mm and cm) grain sizes commonly inferred from the dust opacity index\cite{Kataoka2017,YangLi2020,Lin2023MNRAS.520.1210L}, but it is beyond the scope of this paper to address the tension in detail.

The inclusion of scattering greatly complicates the radiative transfer equation\cite{Y16}. While Monte Carlo techniques have proven effective in handling scattering in the 3D structure of discs, dealing with aligned grains in a complete manner is notoriously challenging, and much of the research in this area has been limited to spherical or randomly aligned grains\cite{W02, D12, S13, W13, B19, R16, P17}. 
The radiative transfer simplifies significantly if each local patch of the dust disk is approximated by a plane-parallel slab\cite{Lin2022MNRAS.512.3922L}. The plane-parallel approximation is reasonable because the dust layer responsible for the (sub)millimeter continuum emission of HL Tau is geometrically very thin vertically, with a dust scale height of approximately 1 au at a radius of 100 au, based on the lack of an azimuthal variation of the gap width\cite{Pinte2016}.


The dust opacities and the scattering properties depend on the composition and the shape and size of the grains. The grain shape is uncertain and expected to be at least somewhat irregular. To facilitate quantitative modeling, we represent the grains with effective spheroids\cite{B83}, whose scattering properties can be calculated using the T-matrix method\cite{M96a, M96b}. To be explicit, when we refer to ``prolate" grains, we refer to grains that are effectively prolate (including intrinsically elongated grains and grains spinning around their longest axis) and likewise ``oblate grains" for effectively oblate grains (including intrinsically oblate grains and grains spinning around their shortest axis)\cite{Yang2019}. Since there is already evidence that polarized thermal emission from effective prolate grains can explain the azimuthal polarization pattern in the HL Tau disk at Band 3 (3.1\,mm) better than that from effective oblate grains\cite{Yang2019,M21,Lin2022MNRAS.512.3922L}, we choose to use prolate grains to model the similarly azimuthal pattern observed in the gaps in our high-resolution Band 7 data as well. As we will see in the Extended Data Fig.~\ref{fig:model_I_lpi_lpol} (panel a), they naturally produce the parallel-to-the-major-axis polarization orientations observed in the first gap at the locations along the minor axis. This important feature is difficult, if not impossible, to produce with effectively oblate grains, which will be shown later in this section. The polarization fraction of the thermal radiation emitted by aligned prolate grains depends on the grain's alignment efficiency and intrinsic degree of elongation, which are uncertain. In particular, the former can depend on the grain size, especially if the grains are aligned by a mechanism depending on internal alignment, which is size-dependent\cite{H09}. In our model, these effects are encapsulated in a dimensionless free parameter, $s$, the effective aspect ratio of the adopted prolate grain. For simplicity, we assume the same value of $s$ for grains of all sizes in the power-law size distribution, with the expectation that the value inferred from model comparison with observations is indicative of the effective aspect ratio of mainly those grains contributing most to the thermal emission -- the largest grains for the adopted size distribution.

Obtaining the dust continuum intensity from the plane-parallel slab requires the alignment direction of prolate grains, temperature $T$, and surface density $\Sigma$ at each line-of-sight. For simplicity, we assume the disk is axisymmetric, and we prescribe $T$ and $\Sigma$ as a function of the cylindrical radius $R$. As motivated above, we consider only toroidally aligned, prolate grains that do not depend on the radius. To limit the parameter space, we use a power-law prescription for the temperature profile
\begin{align}
    T = T_{10} (R / 10\, \text{au} )^{-0.5} ,
\end{align}
where $T_{10}$ is the temperature at a radius of 10~au. We fix the power-law index to $-0.5$ since it is the expectation of a passively irradiated disk\cite{C97,O22} and is consistent with previous constraints\cite{C19}. 

As an approximation to the set of multiple rings, we construct the dust surface density distribution as a summation of individual axisymmetric Gaussian rings. The surface density of the $i$-th ring is defined by its absorption optical depth through 
\begin{align}
    \tau_{i}((R) = \tau_{c,i} \exp{ \bigg( - \frac{1}{2} ( \frac{R - C_{i}}{W_{i}} )^{2} \bigg) } ,
\end{align}
where $C_{i}$ is the location of the center of the ring, $W_{i}$ is the scale width, and $\tau_{c,i}$ is the optical depth at the ring center. The surface density is related to the optical depth through $\Sigma_{i}(R) = \tau_{c,i} / \kappa_{\text{ext}}$ where $\kappa_{\text{ext}}$ is the dust extinction opacity. 
Since the grains are non-spherical, the extinction opacity should depend on the viewing direction of the grain, but here we take $\kappa_{\text{ext}}$ as the extinction opacity averaged over all solid angle.

In a similar manner, the inner disk (denoted by a subscript $a$) is prescribed as a power-law with an exponential taper also defined through the optical depth by
\begin{align}
    \tau_{a}(R) = \tau_{c,0} \bigg( \dfrac{R}{ R_{0} } \bigg)^{-p} \exp{ \bigg( \dfrac{R}{R_{0}} \bigg)^{1.5} } ,
\end{align}
where $\tau_{c,0}$ is the characteristic optical depth and $R_{0}$ is the characteristic radius. The surface density of the complete disk is simply a summation of the rings and the inner disk expressed as:
\begin{align}
    \Sigma(R) = \Sigma_{a}(R) + \sum_{i=1}^{N} \Sigma_{i}(R) ,
\end{align}
where $N$ is the total number of rings.

Following the methodology from Lin et al. (2022), creating the complete image of the disk involves piecing together each patch of the disk whose intensity is approximated by the emergent intensity of the plane-parallel slab. First, we produce the image of the disk in the ``principal frame" where the image coordinates, $x_{p}$ and $y_{p}$, are along the disk minor and major axes direction given some inclination $i$ ($i=0$ means face-on). The Stokes convention of ALMA is based on the IAU (1974) convention\cite{H96}. For clarity, let $Q_{p}$ and $U_{p}$ be the Stokes parameters $Q$ and $U$ in the principal frame with the IAU (1974) convention. The disk inclination produces polarization parallel to the disk minor axis (along the $x_{p}$ direction) which means a positive Stokes $Q_{p}$. Second, since the observations are defined in RA and Dec (i.e., ``sky frame"), we convert the Stokes parameters $Q$ and $U$ in the principal frame (i.e., $Q_{p}$ and $U_{p}$) to those in the sky frame (denoted by subscript $s$, i.e., $Q_{s}$ and $U_{s}$) through
\begin{align} \label{eq:rotate_QU_principal_to_sky}
    Q_{s} &= Q_{p} \cos 2 \eta - U_{p} \sin 2 \eta \\
    U_{s} &= Q_{p} \sin 2 \eta + U_{p} \cos 2 \eta ,
\end{align}
where $\eta$ is the position angle (counterclockwise from North) of the disk minor axis on the far side. Stokes~$I$ in the principal frame and that in the sky frame (Stokes~$I_{s}$) are equal in value. From previous observations\cite{ALMA2015}, we adopt an inclination of $i=46.7^{\circ}$ and a position angle of the disk major axis of $138.02^{\circ}$. Given that the blueshifted outflow is to the northeast\cite{ALMA2015}, the position angle of the far side of the disk is at $\eta=48.02^{\circ}$. Lastly, when comparing the model to the observations of finite resolution, we convolve the model with the elliptical Gaussian beam of the observation.

We try to match the Stokes~$I_{s}$ and the linear polarization fraction along the disk major and minor axes. The Stokes~$I_{s}$ profile constrains the locations and widths of the rings, the temperature $T_{10}$, and in part, the optical depth. The polarization fraction constrains $a_{\text{max}}$, $s$, and the optical depth\cite{Yang2017}. Extended Data Table~\ref{tab:model_parameters} shows the adopted parameters. In general, each bright band identified from the observations can be reproduced using a single Gaussian ring. We find that $a_{\text{max}}=100$~$\mu$m matches well with the observations, and such a fit is similar to previous results based on polarization\cite{K16, M21}. As a reference, the adopted $a_{\text{max}}$ gives $\kappa_{\text{ext}}\sim3.12$~cm$^{2}$ g$^{-1}$ which provides the conversion from the optical depth to the dust surface density. Although we find that the adopted parameters appear adequate enough to broadly reproduce the observations, we caution that the parameters may not be unique in producing such features\cite{Tazaki2019ApJ...885...52T,YangLi2020,Lin2023MNRAS.520.1210L}.

Extended Data Fig.~\ref{fig:model_I_lpi_lpol}a shows the resulting model for the Stokes~$I$ and polarization vectors that can be compared to Fig.~\ref{fig:polmorph}. Extended Data Figs.~\ref{fig:model_I_lpi_lpol}b and~c show the polarization fraction and polarized intensity that can be compared to Fig.~\ref{fig:polimages}. The close similarity between the model and the observed image demonstrates that scattering by aligned prolate grains together with their polarized thermal emission and extinction can readily capture the main observed features, as discussed in the main text. 
First, the polarization fractions and polarized intensities (Extended Data Fig.~\ref{fig:model_I_lpi_lpol}b,~c) are mostly larger along the major axis. The origin is because the thermal polarization and inclination-induced polarization are constructive along the major axis, but destructive along the minor axis\cite{Lin2022MNRAS.512.3922L}. In addition, beam averaging of the rings and gaps also contributes to this azimuthal variation. The polarization from the rings that are uniformly parallel to the disk minor axis are constructive with the thermal polarization from the gaps along the major axis, but destructive along the minor axis.

Second, along the major axis, the polarization fraction is much higher in the gaps than in the rings, which we can easily identify for the innermost five gaps (D14, D34, D44, D68, and D78) from Extended Data Fig.~\ref{fig:model_I_lpi_lpol}b. This is also the case for polarized intensity though only easily seen at D34 and D44 from Extended Data Fig.~\ref{fig:model_I_lpi_lpol}c. To show the trend more clearly, we show the one dimensional profile along the major in Extended Data Fig.~\ref{fig:model_major}, and compare the model to the observation. We also do this for the minor axis in Extended Data Figure~\ref{fig:model_minor}. To show the ``direction" of the polarization fraction, we also plot $q_{p} \equiv Q_{p} / I_{p}$ in Extended Data Figure~\ref{fig:model_minor}. Since the polarization along the principal axes is either parallel or perpendicular to the minor axis direction for the disk model, using $q_{p}$ alone is sufficient and its absolute value corresponds to $p$. A positive $q_{p}$ means polarization parallel to the disk minor axis, while a negative $q_{p}$ means polarization parallel to the disk major axis. For the observation, we calculate its $Q_{p}$ through Eq.~(\ref{eq:rotate_QU_principal_to_sky}) given the observed $Q_{s}$ and $U_{s}$. The model is able to capture the distribution of the Stokes~$I$ and its anti-correlation with the polarization fraction.

Qualitatively, the anti-correlation is a natural result of the varying optical depth of the rings and gaps. Thermal polarization is most prominent when the emitting material is optically thin. However, when the optical depth increases, dichroic extinction from the same material causes the thermal polarization to drop\cite{H00, Yang2017, Lin2022MNRAS.512.3922L}. In contrast, scattering, and thus polarization due to scattering, only becomes effective when the optical depth reaches order of unity or larger\cite{Yang2017}. Therefore, the optically thick rings can have large Stokes~$I$ but low $p$, while the optically thinner gaps can have low Stokes~$I$ and large $p$.

For quantitative comparisons, we also show the cuts of the model image before convolution. The units of Stokes~$IQU$ are scaled by the same factor as the convolved model for easier comparison. Convolution averages the intensities of the rings and gaps, but the polarization from the gaps are more affected than the polarization from rings. Prior to convolution, the polarization fraction in the gaps is much higher than the polarization fraction after convolution. For example, at the first gap (D14), the polarization is $\sim 7\%$ before convolution, but appears as $\sim 4.5\%$ after convolution.

Furthermore, the thermal polarization from toroidally aligned prolate grains can explain the polarization in the minor axis of the first gap (D14) which is \textit{perpendicular} to the disk minor axis (opposite to the inclination-induced polarization). From Extended Data Fig.~\ref{fig:model_minor}, we show the cut along the minor axis. As defined above, the positive offset is along the far side of the disk. The convolved model is able to capture similar levels of the polarization fraction in the inner disk, the first gap (D14), and the ring between D14 and D34 (the first ring). Effects of beam averaging are more prominent along the minor axis since the gap is narrower in projection. Prior to convolution, the polarization fraction in D14 is $\sim 5\%$, but appears as $\sim 1.7\%$. However, at the inner disk, the peak polarized intensity and the peak polarization fraction are not at the center as predicted from the model, but shifted to the near side (southwest; Fig.~\ref{fig:polimages}). The shift can at least in part be explained by the near-far side asymmetry of a geometrically thick inner dust disk\cite{Yang2017}. The model cannot capture the near-far side asymmetry since the image is built from plane-parallel slabs that is more applicable for geometrically thin disks.

We find that the aspect ratio of the prolate grain is $s=0.85$ if perfectly aligned. Larger values of $s$ make the polarization in the gap too strong, while smaller values of $s$ make the elliptical polarization pattern disappear, particularly along the minor axis. For convenience, consider $p_{0}$ as the intrinsic level of polarization of the grain viewed edge-on (with the axis of symmetry perpendicular to the line-of-sight) in the optically thin limit (see Ref.~\citenum{Lin2022MNRAS.512.3922L} for details). The prolate grain with $s=0.85$ gives $p_{0} \sim 12\%$ which is comparable to the observed levels of polarization in the envelope scale\cite{C15, C18, K19, L20} and in the interstellar medium\cite{Planck19,Planck20}. The high $p_{0}$ is necessary because it works against the optical depth of the gap itself and contamination from rings due to beam averaging.

In contrast, the outer regions along the minor axis do not show similar polarization perpendicular to the disk minor axis. As demonstrated in Extended Data Fig.~\ref{fig:model_major}, this can be explained by beam averaging of several rings with large enough optical depth. Going to even higher angular resolution and better sensitivity should mitigate the beam averaging effects and uncover the thermal polarization from aligned grains if any. Such an observation would make measurements of the radial distribution of grain alignment possible and provide constraints on the alignment mechanism.

The aspect ratio of $s=0.85$ is much smaller than the $s\sim0.97$ inferred from the lower angular resolution image of Band~3 (3.1 mm) of HL Tau\cite{Lin2022MNRAS.512.3922L}. The latter is constrained from the $\sim 2\%$ polarization\cite{Kataoka2017,Stephens2017c} along the outer regions of the disk's minor axis in Band~3 when assuming a smooth disk. The contrast suggests that polarization from aligned grains are diluted by polarization from the optically thick substructures even at Band~3. Another possibility is from varying levels of grain alignment efficiency or aspect ratios. The inner region may be more aligned/elongated, while the outer regions are less aligned/elongated. Going to longer wavelengths with similar angular resolution to resolve the rings and gaps will help distinguish the two scenarios. The lower optical depth at longer wavelengths increases the polarization fraction from aligned grains in the gaps, and also the decreased optical depth of the rings effectively increases the angular extent of the optically thin gaps making it easier to resolve the gaps. 

If the dilution of large $p_{0}$ grains from optically thick substructure is true, then it has far reaching implications for detecting grain alignment in disks in general. ALMA has routinely detected polarization for disks though with only a few beams across the disk minor axis. Many sources show an elliptically-oriented polarization pattern of a few percent in the outer regions\cite{Ba18, Harrison2019, M19, S19, H21} which is similar to the previous low angular resolution observations of HL Tau at Band~3\cite{Stephens2017c}. Since most of these sources also have substructure, we speculate that the intrinsic polarization should be much higher just like the case of HL Tau and grain alignment may be frequent. More high angular resolution studies of various sources that can resolve the gaps will be necessary to verify the frequency of aligned grains and its prevalence as a general phenomena in disks.


We also rerun this identical analysis, except we use oblate grains with an aspect ratio of $s = 1.2$ and their short axes aligned radially in the disk. We do not consider oblate grains that are toroidally aligned, because the polarization would be radial and completely different from what is observed. The results are shown in Extended Data Fig.~\ref{fig:oblate}. In the bottom panels (panels a and b) of this figure, we saturate the color-scale for the polarization fraction and morphology so that one can directly compare to Extended Data Fig.~\ref{fig:model_I_lpi_lpol} for the prolate grain model. This model is inconsistent with the observations since the polarization direction is opposite to what is observed in D14 along the minor axis. This discrepancy in polarization orientation can be removed in principle by increasing the aspect ratio of the oblate grain, but it would make the already excessively high polarization at locations along the major axis even higher. Decreasing the aspect ratio would render scattering more dominant, worsening the polarization orientation discrepancy in D14 along the minor axis and causing the polarization orientations in the gaps less azimuthal than observed.

To a certain extent, pure scattering of spherical or randomly aligned grains with the radiation anisotropy due to the existence of rings and gaps can also produce the anti-correlation between Stokes~$I$ and $p$ along the major axis. Namely, the radiation anisotropy is greater in the gaps than in the rings, which leads to larger $p$ (Ref. \citenum{D19, Ohashi2019ApJ...886..103O, Lin2020b}). However, solely relying on radiation anisotropy will be difficult to reproduce both the polarization direction and the high level of polarized intensity along the minor axis for the first gap (D14). Although each scattered photon can be highly polarized, one needs material in the gaps to scatter enough photons from the rings in order to produce the high polarized intensity, but the polarization tends to follow the disk minor axis (inclination-induced polarization) with increasing optical depth.



Radiation anisotropy is not included given the plane-parallel nature of the model, and it is unclear how it will affect the inferred $s$ (also degenerate with the alignment efficiency). As mentioned above, given the large optical depth seen along the midplane between each neighboring patch of the disk, we can expect marginal contributions from radiation anisotropy which makes the use of plane-parallel slab calculations favorable. This is likely the case for gaps too since the optical depth to the observer in the gaps is already greater than 1 (Extended Data Figs.~\ref{fig:model_major} and~\ref{fig:model_minor}). However, future refinements of the model will need to consider radiation anisotropy in 3D geometry. The strongest constraint for $s$ comes from the polarization in the first gap (D14), where the radiation anisotropy should be larger compared to the rings. The 3D geometry is also necessary to explain the near/far side asymmetry of the inner disk. Monte Carlo radiation transfer codes that can easily handle radiation anisotropy in 3D geometry do not typically include scattering of aligned grains\cite{D12}. The incredible quality of the data highlights the necessity for the development of 3D radiation transfer codes to include thermal emission and scattering of aligned grains in a self-consistent manner.



\end{methods}

\newpage
\renewcommand{\figurename}{\textbf{Extended Data Fig.}}
\setcounter{figure}{0}
\renewcommand{\tablename}{\textbf{Extended Data Table}}
\setcounter{table}{0}
\begin{table}
    \centering
    \begin{tabular}{c|ccccccccccc}
        $T_{10}$ [K] & 110 \\
        $a_{\text{max}}$ [$\mu$m] & 100 \\
        $s$ & 0.85 \\
        $R_{0}$ [au] & 10   \\
        $\tau_{a}$ & 2.3 \\
        $p$ & 0.5 \\
        $C$ [au]& 24 & 39 & 49 & 59 & 73 & 88 & 102 & 116 \\
        $\tau_{c}$ & 8 & 5 & 5 & 8 & 3 & 8 & 8 & 8  \\
        $W$ [au] & 4 & 3 & 3 & 2 & 4 & 3 & 2 & 2 \\
    \end{tabular}

    \caption{
        The parameters used to produce the disk model and polarization image. 
    }
     \label{tab:model_parameters}
\end{table}

\begin{figure}
    \centering
\includegraphics[width=0.5\textwidth]{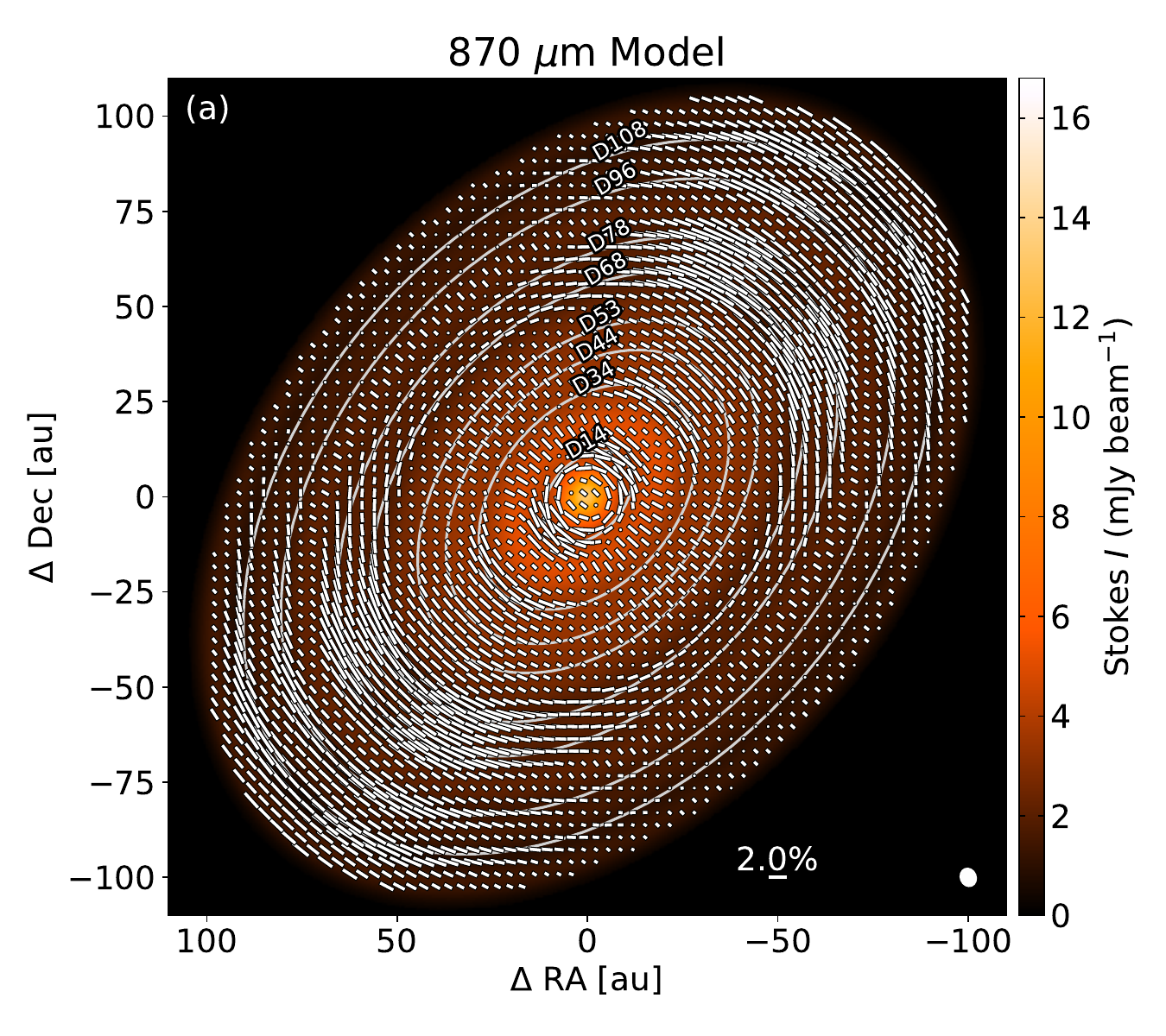}
\includegraphics[width=0.9\textwidth]{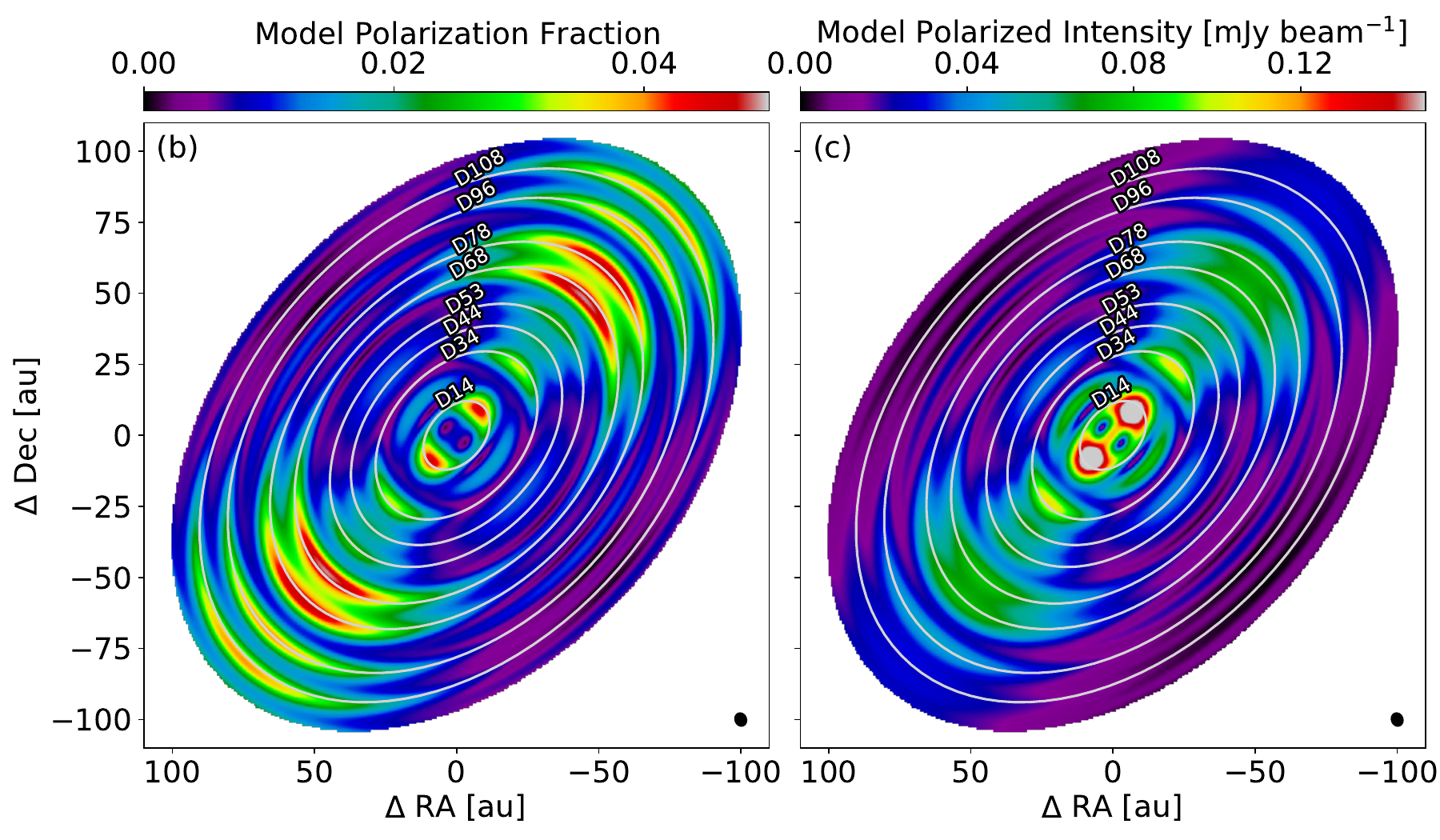}
    \caption{
        {\bf Polarization images of the model.} {\bf a,} The colormap is the Stokes~$I$ image in mJy\,beam$^{-1}$ and the line segments represent the polarization angle. The length of the segments are proportional to the polarization fraction with a $2\%$ scale bar shown in the bottom. {\bf b,} The linear polarization fraction image. {\bf c,} The linear polarized intensity in mJy\,beam$^{-1}$. The resolution is shown as a small white ellipse in the bottom right of each panel. The concentric ellipses on top of the disk mark the location of the gaps. 
    }
    \label{fig:model_I_lpi_lpol}
\end{figure}

\begin{figure}
    \centering
    \includegraphics[width=\textwidth]{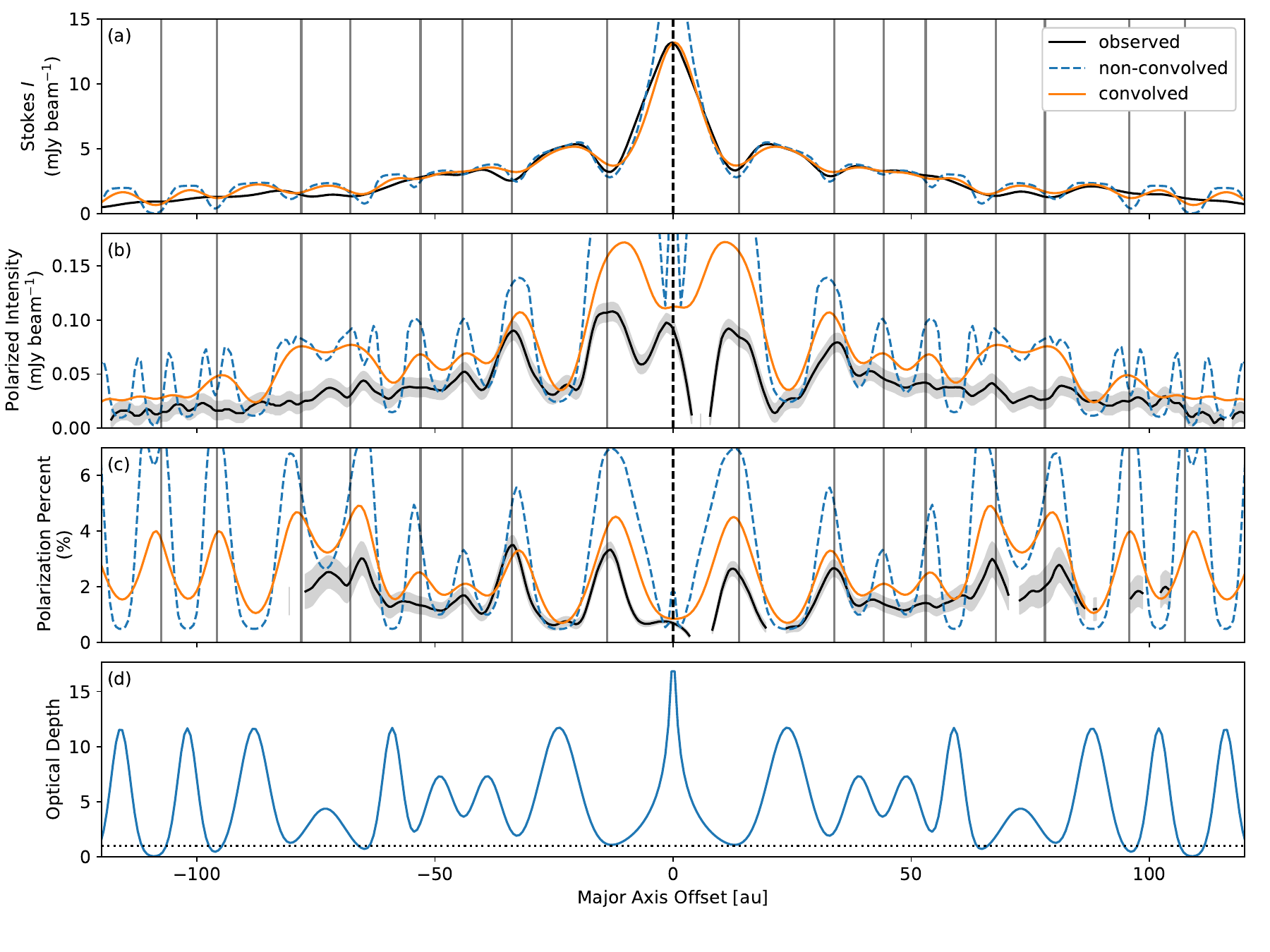}
    \caption{
        {\bf Comparisons of the profiles along the major axis between the observation and model.} The first three panels show the Stokes~$I$ (panel a), linear polarized intensity (panel b), and the linear polarization fraction (panel c). The observations are plotted in black solid lines with shaded areas showing the standard deviation. The dashed blue lines show the model profiles prior to beam convolution and the orange solid lines show the model profiles after beam convolution. Panel d shows the input optical depth of the model. The horizontal dotted line is where the optical depth is 1. The vertical solid lines mark the locations of the gaps. 
    }
    \label{fig:model_major}
\end{figure}

\begin{figure}
    \centering
    \includegraphics[width=\textwidth]{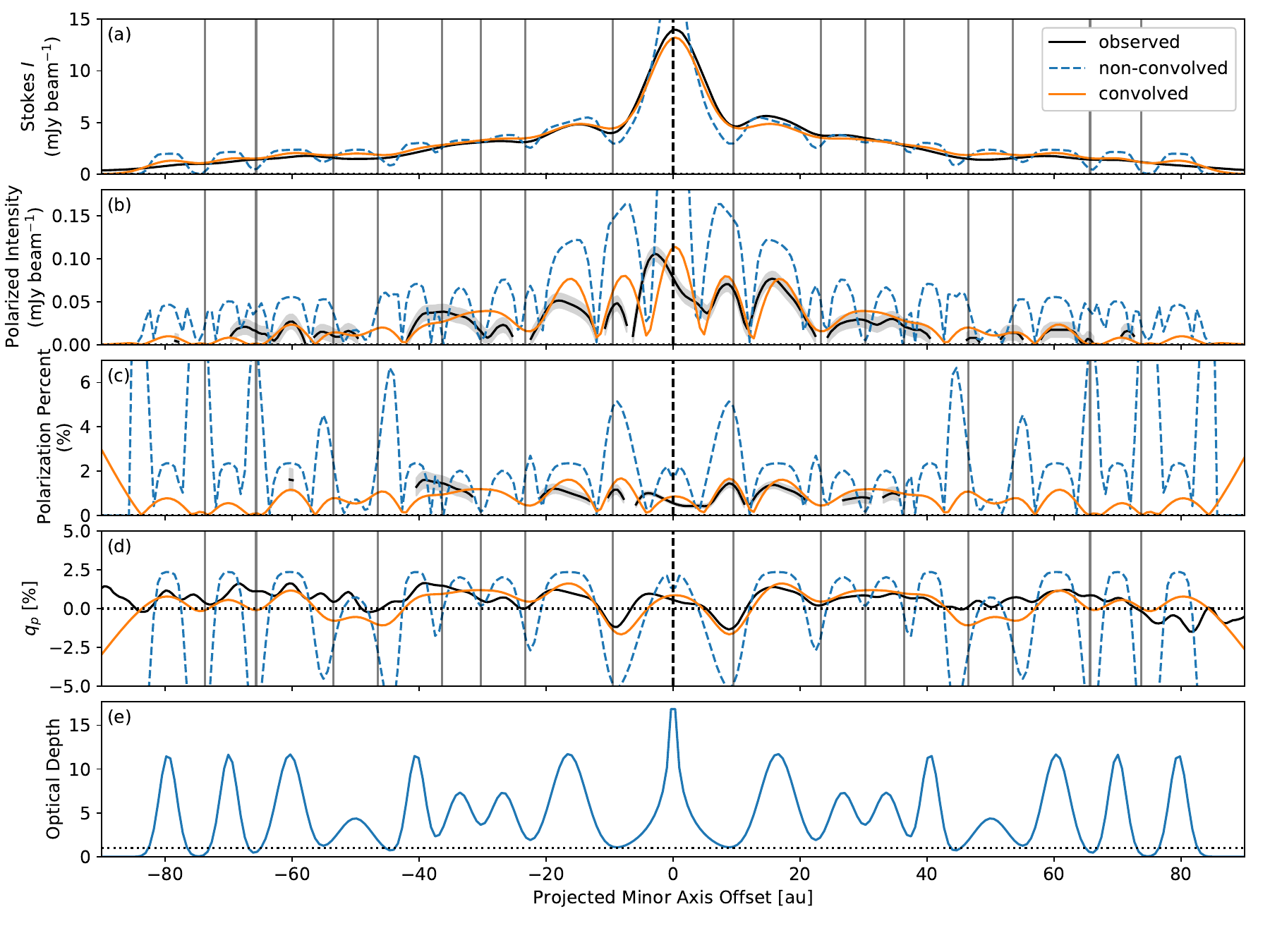}
    \caption{
        {\bf Profiles along the minor axis.} These are plotted in a similar way as Extended Data~\ref{fig:model_major}. The additional panel d shows $q_{p}$ (see text for the definition). The positive offset is towards the northeast direction. 
    }
    \label{fig:model_minor}
\end{figure}

\begin{figure}
    \centering
\includegraphics[width=0.5\textwidth]{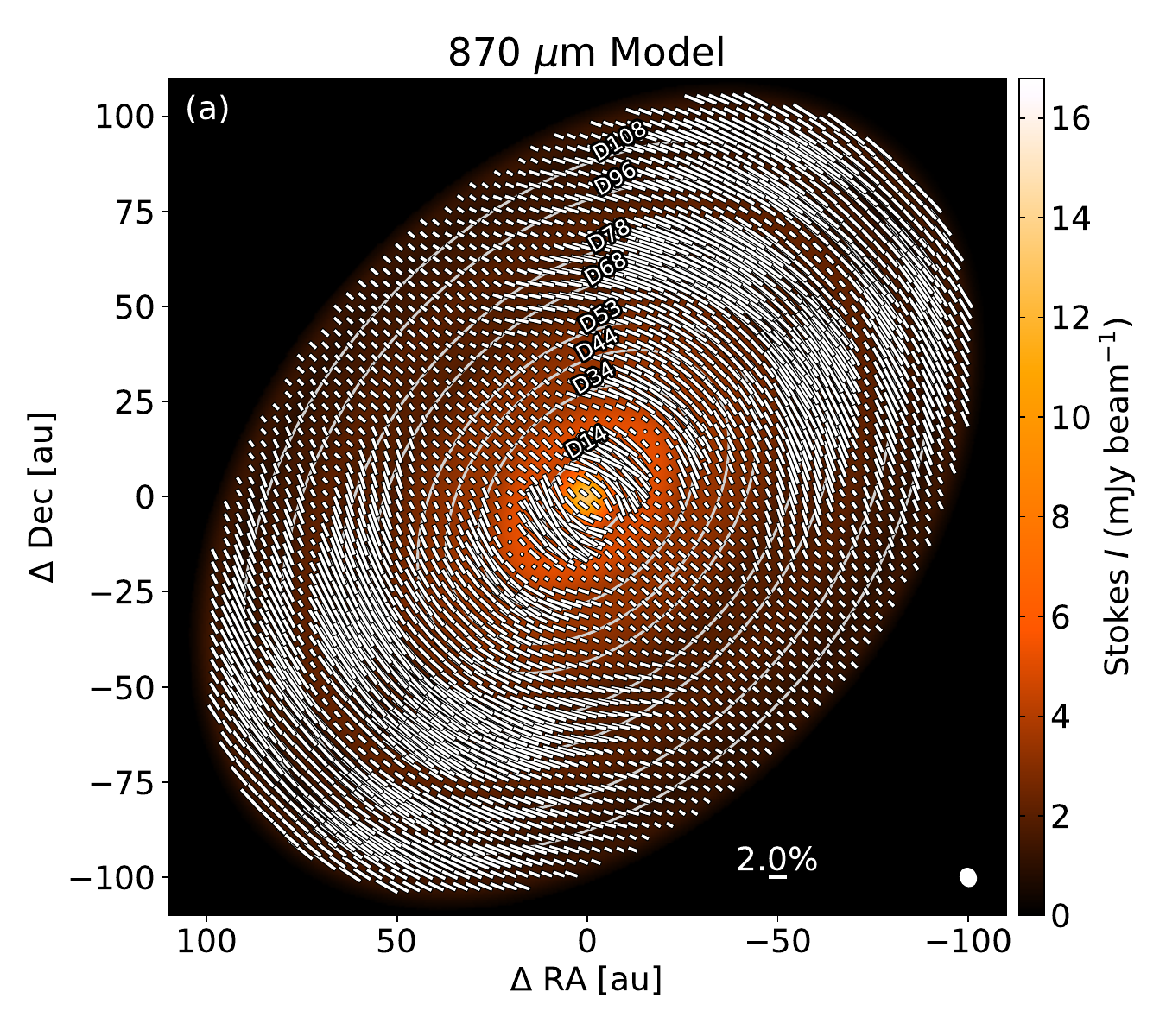}
\includegraphics[width=0.9\textwidth]{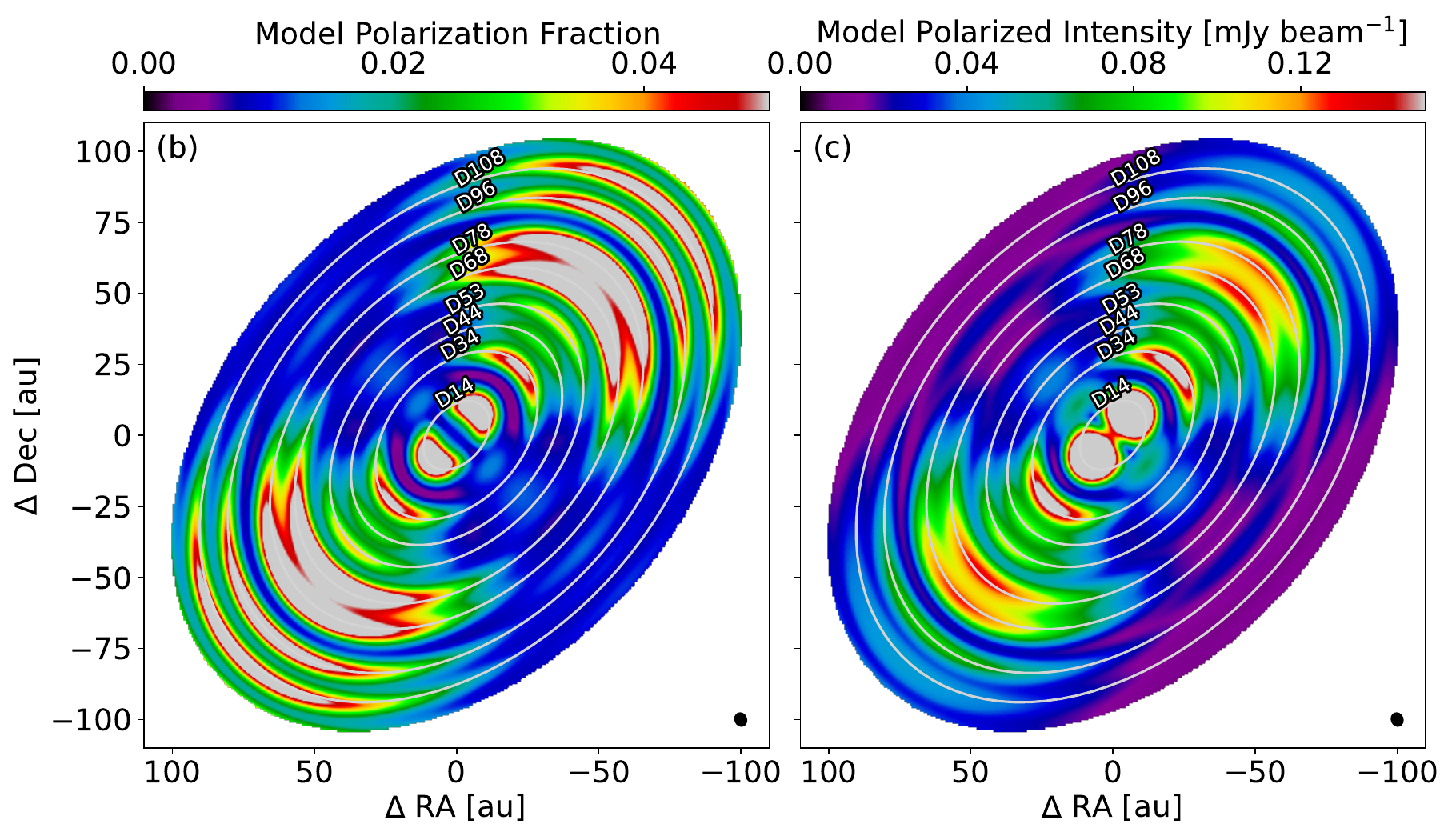}
    \caption{
       {\bf Model using effectively oblate grains.} Figure caption is the same as Extended Data Fig.~\ref{fig:model_I_lpi_lpol}, but now with oblate grains. The color scales in panels b and c have been saturated at many locations so that the morphologies of the low-level polarization fractions and intensities are visible.
    }
    \label{fig:oblate}
\end{figure}

\end{document}